\documentstyle[12pt] {article}
\begin {document}
\parindent=15pt
\begin{center}
\vskip 1.5 truecm
{\bf RAPIDITY DISTRIBUTIONS OF JETS PRODUCED IN HEAVY ION COLLISIONS
AT LHC}\\
\vspace{.5cm}
G. Calucci and D.Treleani \\
\vspace{.5cm}
Dipartimento di Fisica Teorica dell'Universit\`a and INFN.\\
Trieste, I 34014 Italy
\end{center}
\vspace{1cm}
\begin{abstract}
We discuss the rapidity distribution of produced jets in heavy-ion
collisions at LHC. The process allows one to determine to a good accuracy
the value of the impact parameter of the nuclear collision in each single
inelastic event. The knowledge of the geometry is a
powerful tool for a detailed analysis of the process, making it possible
to test the various different elements which, in accordance with 
present theoretical ideas, take part to the production mechanism.
\end{abstract}
\vspace{3cm}

\vspace{3cm}

E-mail giorgio@trieste.infn.it \\

E-mail daniel@trieste.infn.it \\

\newpage

\section{Introduction}

Within a few years it will be possible to observe at LHC the first events 
with heavy ions colliding at energies of several TeV in the nucleon-nucleon 
c.m. system. The interest in the process is justified by the expectation that 
a new phase of matter, namely
quark-gluon plasma, could appear as a result of the large energies - high
densities which will be reached accelerating heavy ions at LHC 
energies \cite{Q-G}.
On the other hand the large number of fragments, resulting from such large
structures, as say Pb nuclei, interacting with so large energy in the c.m.
system, makes the
analysis of the events rather difficult and the search of unambiguous
signatures puzzling. The identification of a few features in the event which 
can be easily understood and whose dynamics is, on the contrary, rather
transparent would be in this respect an important handle. The production
of jets is, in our opinion, a process which has good possibilities to
be understood in a rather satisfactory way, also at such an extreme regime
as in the case of heavy ion collisions at LHC \cite{jets}. 
In fact, in the present paper, we discuss
the mechanism of production of large $k_t$ jets and of minijets 
in heavy ion collisions at LHC energies. The main
observation is that one has the possibility of disentangling the geometrical
features and to obtain as a consequence
distributions with universal characteristics. When geometry is
decoupled one has in fact the possibility of constructing various
distributions of jets, produced 
at different rapidities, with different cut-off values etc.,
which are related one to another in a simple way,
that is easily worked out in perturbation theory. This universality property,
on the other hand, represents a regularity feature which is 
easily tested experimentally and which is rather  
independent on the details of the actual theoretical approach. 

\section{Discussion}
The inclusive cross section to produce jets at large $k_t$ is expressed, in the
QCD-parton model, by the factorized form
\begin{equation}
{d\sigma_J\over dx_Adx_Bd^2k_t}=G(x_A)G(x_B){d\hat{\sigma}\over d^2k_t}
\end{equation}
where $G(x)$ is the parton distribution, depending on the fractional momentum
$x$ and on the scale $k_t$, and $d\hat{\sigma}\over d^2k_t$ is the partonic
cross section. By introducing a lower cutoff in the transverse momentum of
the produced jets $k_t^c$ one defines the integrated cross section:
\begin{equation}
\sigma_J(k_t^c)=
\int_{sx_Ax_B>4(k_t^c)^2}{d\sigma_J\over dx_Adx_Bd^2k_t}dx_Adx_Bd^2k_t
\end{equation}  
The lower limit for the integration in $x_A$ and $x_B$ 
is obtained by the condition
$sx_Ax_B>4(k_t^c)^2$. The differential cross section
as a function of $x_A$ (or as a function of $x_B$) 
results from the corresponding integration 
on $x_B$ (or on $x_A$). For $x_A$ one has to integrate $x_B$ 
with the limits  $1>x_B>4(k_t^c)^2/sx_A$.
The expression of the differential cross section 
as a function of $x_A$ is therefore written as
\begin{equation}
{d\sigma_J\over dx_A}=G(x_A)\times
\int_{x_B>4(k_t^c)^2/sx_A}G(x_B){d\hat{\sigma}_J\over d^2k_t}dx_Bd^2k_t
\end{equation} 
The relation corresponds to a flux of projectile partons, at a given value
of momentum fraction $x_A$, interacting with the target. The integral
represents the target scattering centers which are seen by a projectile
parton with momentum fraction $x_A$. One may notice,
by looking at the integration limits of the integral on $x_B$,
that when $x_A$ is large the
number of scattering centers is also large, while at small $x_A$ the 
number of scattering centres drops to zero. The size of each scattering
centre is set by the value of the partonic cross section, and, although
it depends on the value of the cut-off $k_t^c$, for sensible values
of $k_t^c$, i.e. for values of the order of a few $GeV$, 
the size of the scattering centres 
is very small as compared to the hadronic dimension.

When considering large nuclei interacting at LHC energy the number of scattering
centres which is effective for producing jets with momentum transfer 
larger than a few $GeV$ is a rather large number. The scattering centres are 
distributed in the volume occupied by the target nucleus. One may 
figure out the interaction process by integrating the hadronic parton
distributions with the limits discussed above and by multiplying the result
by the atomic mass number. The resulting quantity is the number of scattering
centres as a function of the momentum fraction of the projectile $x_A$.
One may then distribute the scattering centres uniformly in the volume of
the target nucleus and take the projection in the transverse plane. In
correspondence of each scattering centre, projected in the transverse plane,
one can draw a dot whose size is the parton-parton interaction cross
section integrated with $k_t>k_t^c$. The result is shown in fig.1
for collisions with $7TeV$ in the nucleon-nucleon c.m. system, $5GeV$
as a value of $k_t^c$ and $Pb$ as a nuclear target. The two figures
show the distribution of scattering centres for two different values of
$x_A$, $x_A=10^{-3}$ and $x_A=10^{-1}$, corresponding to the case of
mini-jets produced in the central and in the forward rapidity region.
The scale in fig.1 is the $fm$ and the size of the dot representing each
scattering centre has been obtained by computing the partonic cross section
in Born approximation. In the actual forward rapidity case
the number of possible
scattering centres which are effective is of the order of $20.000$.

Fig.1 points out in a spectacular way one of the most relevant problems
one has to deal with when considering production of jets in high energy
nuclear collisions. Namely the large number of elementary interactions
which take place in a single inelastic event and the geometrical features
which result from the localization of the elementary interactions.
The first observation is that when the number of
scattering centres is very large, there may be overlaps in a few cases. As a 
consequence the effective number of scattering centres is smaller with respect
to the number that has been estimated simply by multiplying the parton
distributions with the atomic mass number. To keep this feature into account one
should consider the {\it gluon-fusion} process \cite{gfus}, which reduces 
the parton population accordingly. One can then observe that the most 
peculiar feature of the overall interaction is represented by the large number
of elementary partonic collisions, which take place at different points in 
the transverse plane, inside the overlap region of the matter distribution 
of the two interacting nuclei. One may in fact expect that the parton population
of the projectile is also distributed in transverse space in a way similar to
the distribution of the scattering centres shown in fig.1. The hard, or
semi-hard, nuclear collision is then obtained by superposing two pictures
similar to those shown in fig.1, one picture representing the distribution
of projectile partons and one representing the distribution of target
partons. Both projectile and target partons are localized in transverse plane
in a region whose size is set by the cut-off $k_t^c$ and which is of the
order of the size of the partonic interaction
cross section, as in the case of the dots in fig.1. 
Projectile and target partons
interact when the regions where they are localized in transverse plane
overlap. The interactions are then localized in the overlap region of
the matter distribution of the 
two nuclei and the whole process depends strongly on
the nuclear impact parameter
as a consequence. 

There are two qualitatively different
kinds of multiple partonic interactions:
\begin{itemize}
\item different pairs of partons interact at different
points in transverse plane, 
that one may call {\it disconnected interactions} \cite{disconn}, and 
\item a given projectile parton may overlap at the same time with two
or more target partons, 
that corresponds to a {\it rescatterings} \cite{resc} process. 
In this respect one may notice
that there is no need to be in the regime where parton fusion is an important
effect in order to have a sizeable rescattering probability.
\end{itemize}
The final state produced by a single 
elementary partonic interaction is represented by two 
or more scattered partons. The simplest possibility is to have just 
a Born level process in such a way that one has a one to one correspondence 
between initial state partons and jets observed in the final state.
Further jets produced in the single hard interaction are then to be regarded as 
corrections of order $\alpha_S$ to the lowest order process. 

The quantities which are needed for the actual description of the nuclear hard 
interaction process are 
the (non-perturbative) initial state partonic nuclear distribution and the 
(perturbative) interaction
probability. Because of the multiple partonic interactions,
the information which is needed on 
the initial state is however much more detailed
with respect to the nuclear parton distributions, usually considered
in inclusive processes. As one realizes looking at fig.1, 
one needs in fact to know not only the number of partons
with given momentum fraction but also their position in transverse 
plane. In addition, given the number of partons with given momentum
fraction and at a given position in transverse plane, one needs to know 
the parton population at a different transverse point and with 
a different $x$ value. The non-perturbative input to the process is
in fact represented by the whole nuclear multi-parton
distribution \cite{multiparton}, which corresponds to a 
much more detailed knowledge
on the nuclear parton structure than presently available. One needs
in fact to know the whole set of all multi-parton correlations,
which are independent quantities with respect to the parton distributions
usually considered.

It seems however reasonable that when dealing with many partons
localized in different space-time regions, as in the case of the
parton population of a large nucleus, correlations are not a 
major effect. When correlations are completely neglected
the parton population is described by a
Poisson distribution. The probability to have a configuration
with $n$ partons with coordinates $x_1,{\bf b}_1,\dots x_n,{\bf b}_n$
is therefore given by
\begin{equation}
\Gamma(x_1,{\bf b}_1,\dots x_n,{\bf b}_n)
={D(x_1,{\bf b}_1)\dots D(x_n,{\bf b}_n)\over n!}
e^{-\int D(x,{\bf b})dx d^2b}
\end{equation}
In this simplest case the whole distribution is expressed by
means of the average number of partons with coordinates
$x,{\bf b}$, namely $D(x,{\bf b})$. The connection with the
usual parton distributions is $\int D(x,{\bf b})d^2b=G(x)$,
where $G(x)$ is the nuclear parton distribution usually considered
in large $k_t$ processes. The whole multi-parton distribution is
singular in the infrared region and
the $x-$integration in the normalization factor is done by
keeping into account the cutoff $k_t^c$. 

Given a configuration with $n$ projectile partons and $m$ target partons,
the interaction probability is obtained from the elementary 
probability of interaction
$p_{ij}$, where $i$ and $j$ label a given projectile and a given
target parton respectively.
Then $1-p_{ij}$ is
the probability of no interaction for the pair $i,j$ and
$\prod_{i,j}^{n,m}(1-p_{ij})$ represents the probability
of no interaction for the configuration with $n$ projectile partons and $m$ 
target partons. The quantity which one needs for the cross section is the
probability to have at least one interaction \cite{cm}, namely
$P(n,m)\equiv 1-\prod_{i,j}^{n,m}(1-p_{ij})$, which is in fact equal to the 
sum over all interaction probabilities.

The hard cross section $\sigma_H$ is obtained by integrating on the
impact parameter $\beta$ the probability to have at least one hard interaction.
The probability to have at least one hard 
interaction in the nuclear collision is the result of the sum over all partonic
configurations of the projectile and of the target nuclei, multiplied by
the probability to have at least one hard interaction in the given partonic 
configuration.
The hard cross section is therefore expressed as \cite{wounded} 
\begin{equation}
\sigma_H=\int d^2\beta\sum_{n,m}\Gamma_n\cdot\Gamma_m\cdot P(n,m)
\end{equation}

The relation above is too complicated to be read off and
computed directly, one can obtain a closed
analytic form if one takes the drastic attitude of neglecting
all parton rescatterings. One may however derive a few simple relations out 
of Eq.(5). If one works out the average number of parton collisions
$\langle\nu\rangle$ \cite{wounded}
the result is the single scattering expression of the perturbative QCD parton model:
\begin{equation}
\langle\nu\rangle\times\sigma_H=\int_{sx_Ax_B>4(k_t^c)^2} 
G(x_A)G(x_B){d\hat{\sigma}\over d^2k_t}dx_Adx_bd^2k_t
\end{equation}
The relation shows that a rather general non-trivial 
feature, namely the AGK cancellation,   
which is expected to hold 
as a consequence of the AGK cutting rules \cite{AGK} when multiple interactions are present, 
is implemented explicitly in the actual case.

A quantity of interest is the average number of wounded partons \cite{wounded},
namely partons which have
undergone at least one hard interaction. Those are in fact the initial state
partons that are observed as jets in the final state. In the simplest case, 
when the elementary parton-parton interaction is represented as a two to two
process, all jets observed in the final state are wounded partons.
The average number of wounded partons of the projectile nucleus $A$, 
with a given momentum fraction $x_A$ in a nuclear collision
with impact parameter $\beta$, $W(x_A)$, can be derived from Eq.(5).
The resulting expression is:
\begin{equation}
W(x_A)=\int d^2b D(x_A,{\bf b}-{\bf\beta})\cdot
\Bigl\{1-exp\bigl[-\int D(x_B,{\bf b})\hat{\sigma}(x_A,x_B)dx_B\bigr]\Bigr\}
\end{equation}
$\hat{\sigma}(x_A,x_B)$ is the partonic cross section, integrated on 
the momentum transfer $k_t$ with the cut off $k_t^c$ and $x_B$ is integrated
within the limits $4(k_t^c)^2/sx_A<x_B<1$. 
\\
One may notice that, while the initial state parton distribution is
singular for small $x$ and $k_t^c$ values, the average number of wounded
partons is on the contrary a smooth function in the infrared region. In fact
the expression above results from the product of the average number
of $A$-partons, $D(x_A,{\bf b}-{\bf\beta})$, and the interaction
probability, namely 
$\Bigl\{1-exp\bigl[-\int D(x_B,{\bf b})\hat{\sigma}(x_A,x_B)dx_B\bigr]\Bigr\}$.
At small $x_A$ the number of projectile partons is very large, in
that limit the interaction probability however goes to zero. The 
interaction probability in addition has a finite limit, actually the black disk
limit, when $x_A\to1$ or when $k_t^c\to0$.
\\
A further observation is that, when the interaction probability is small,
if the dependence of the average number of partons on $x$ and on ${\bf b}$ is 
factorized,
the expression for the average number of wounded partons is also factorized
in the same variables. More explicitly, if one writes
\begin{equation}
D(x,{\bf b})=G(x)\tau({\bf b})
\end{equation}
where $\tau({\bf b})$ is the nuclear thickness function normalized to one,
one obtains, in the limit of small interaction probability,
\begin{equation}
W(x_A)=G(x_A)\int G(x_B)\hat{\sigma}(x_A,x_B)dx_B\times
\int d^2b \tau_A({\bf b}-{\bf\beta})\tau_B({\bf b})
\end{equation}
The average number of wounded partons $W(x_A)$, in a nuclear
collision with impact parameter $\beta$, is therefore proportional in this case
to the overlap of the matter distribution of the two interacting
nuclei $\int d^2b \tau_A({\bf b}-{\bf\beta})\tau_B({\bf b})$.

Given the explicit form of the cross section, one may work out
all higher moments of the distribution in the number of
wounded partons. The resulting expression is more and more involved
when higher and higher moments are considered. An important simplification is
however obtained, in the distribution of the wounded partons of the projectile,
if one makes the hypothesis of neglecting
all interactions where different projectile partons hit the same target
parton. The set of events of this kind is rather small when compared with
the whole set of possible events. When this simplification is made
the distribution in the number of wounded partons, at a given value
of $\beta$, turns out to be
a Poissonian and it is therefore determined completely by its average
value. A characteristic feature of the interaction is therefore that
the fluctuation of the distribution in the number of wounded partons
is less and less important, as
compared with the average value, when the average value grows. 
It is precisely this feature that allows one to disentangle the geometrical
aspect of the interaction event by event.

When the distribution of the wounded partons is known one can easily simulate
the production of jets in a collision involving heavy nuclei at LHC energy.
In fig.2 one may see how the number of produced jets is distributed in
rapidity in a central $Pb-Pb$ collision with $7TeV$ energy in the 
nucleon-nucleon c.m. system. In fig.2$a$ one shows the distribution of
jets produced with a transverse momentum larger than $20GeV$, in 
fig.2$b$ the cutoff $k_t^c$ is lowered to $10GeV$ and in fig.2$c$
$k_t^c=5GeV$. In the figures the histograms represent the simulated
event. The continuous line is the average number of wounded partons,
as it results after summing projectile and target wounded partons, as expressed
in Eq.(7) using rapidity as a variable.
The dashed line is
the same average number when using rather the approximate expression in Eq.(9),
namely neglecting semi-hard parton rescatterings. 
\\
One may notice that while the average number grows the fluctuation 
around the average grows much less rapidly. The effect of the unitarization
of the hard parton-nucleus interaction,
namely the difference between the continuous
and the dashed line,
is also shown to be a rather sizable effect 
when $k_t^c=5GeV$. 

Since the fluctuations are relatively less important when the average
numbers are large, one may try to determine the value of the impact
parameter by counting, in a given event, all jets produced above some 
lower cutoff $k_t^c$.
In fig.3 we show the distribution in the total number of jets produced 
in central collision with different cutoff values ($20GeV$ fig.3$a$,
$10GeV$ fig.3$b$ and $7GeV$ fig.3$c$). 

In the case of a $7GeV$ cutoff the fluctuation in the number of
produced minijets is $2.5\%$. The overall number of produced minijets
in a central collision with a $7Gev$ cutoff is however larger than $1500$, 
and the actual possibility of detecting such a large number of minijets
is out of reach. A quantity which is likely to be more accessible
experimentally could be 
the overall energy carried by all the produced minijets. Same numbers
of produced minijets can carry different amounts of energy. The 
relative fluctuation
in energy is therefore larger with respect to the relative fluctuation in 
the number of produced minijets.
Anyhow with a $7GeV$ cutoff the fluctuation in energy turns out to be less than
$4\%$. If one lowers the cutoff to $5GeV$ the fluctuation in the energy
carried by all the produced minijets is close to $2.5\%$. 
A similar variation in the average energy of produced
jets is obtained by varying the value of the impact parameter 
by an amount sizeably smaller than $1fm$. It seems therefore plausible
that the measure of the total energy carried by jets, produced with  
relatively small momentum transfer, will allow the
determination of the value of the impact parameter of the nuclear collision
with an uncertainity smaller than $1$ $fm$.
\\
A possible procedure in classifying events with jets in heavy ion
collisions could be therefore the following:
\\
In each single event
one may measure the total energy of the produced jets as a function of
$k_t^c$. Since the distribution is narrower and narrower when $k_t^c$ is 
smaller and smaller,
the energy carried by the jets with relatively small $k_t^c$ allows 
one to estimate
the value of the impact parameter of the collision.
One may then collect all events with the same impact parameter $\beta$.
At each rapidity value and for a given cutoff $k_t^c$ one has 
therefore a distribution
in the number of observed minijets. One expects that all
these distributions are universal, to a large extent. One expects
in fact that the distributions are the same at different rapidity values
and with different cutoffs, the only difference being in the average values.
The dependence of the average values, as a function of rapidity and
cutoff, is obtained through pQCD.

As an example one may write the expression for the average number of 
produced minijets, at a given rapidity $y$, for all events with
the same impact parameter $\beta$. When $k_t^c$ is not too small,
say it is larger than $10-15GeV$,
semi-hard rescatterings can be neglected and one obtains:
\begin{eqnarray}
{dN\over dy}=\int d^2b \tau_A({\bf b}-{\bf\beta})\tau_B({\bf b})
&\times \Bigl[g_A(y)\int_{y'<y}g_B(y')\hat{\sigma}(y-y',k_t^c)dy'\nonumber\\
&+g_B(y)\int_{y<y'}g_A(y')\hat{\sigma}(y'-y,k_t^c)dy'\Bigr]
\end{eqnarray}
here $g(y)=xG(x)$ and rapidities and fractional momenta are connected
by the usual $2\to2$ kinematics. 
\\
The dependence on $y$ and on the cutoff $k_t^c$, which is described by the term
in square parenthesis, is obtained by the usual pQCD parton model
inputs. Geometry is factorized
and it enters in the first term in Eq.(10) only. All different averages of 
distributions with the same impact parameter are related one to another
according with pQCD as shown in Eq.(10). 
When the impact parameter is changed
the distributions are obtained, from the corresponding
ones measured previously, simply by rescaling the averages with 
the value of the overlap function.

One expects in addition that if one 
plots the relative rates of events with different numbers of minijets,
with $k_t$ larger that some fixed value $k_t^c$ and at the same rapidity,
one obtains a quantity which tests the overlap function,
namely which is sensitive to the actual 
distribution in space of the nuclear 
partonic matter. As an example in fig.4$a$ we plot the
relative rates of events with jets at $y=0$ and $k_t^c=20GeV$
in the case of a Woods-Saxon distribution in space of 
the nuclear partonic matter (continuous histogram)
and in the case of an hard-sphere distribution (dottet histogram).
The difference is much more evident if
one plots the ratio of the two histograms (fig.4$b$).

\section{Conclusions}

A large number of jets are expected in the case of nuclear collisions at LHC.
The global features of the events are understood in terms of the
pQCD parton model, once the geometrical features of the interaction are taken
into account. The picture of the interaction 
just described is easily tested: One may first look at
central collision events only, by taking a veto trigger in the forward 
direction.
All central events are expected to show a universal behavior in the
distributions
of jets produced at different rapidity values and with different cutoff
values. The jets observed are in fact the wounded partons, namely the initial
state partons which have undegone at least one hard interaction.
The expectation is that the different distributions in the number
of wounded partons, and therefore in the number of observed jets, are 
close to a
Poisson distribution when the impact parameter of the nuclear 
collision is fixed. 
At fixed impact parameter all different distributions in the number of
observed jets are therefore 
equal. The only difference  
is in the average value, which is calculation in the pQCD parton model
in a staightforward way.

One can then proceed estimating
the impact parameter event by event. That can be done by looking at the 
total energy carried by the jets produced in each single event at small $k_t^c$.
The value of the energy, compared with the corresponding value
in the case of a central collisions, gives the estimate of the size of
overlap of the two interacting nuclei. Within the simple approach
to the nuclear interaction actually described,
collecting all events with the same overlap one obtains
distributions which have to be the same as those observed in 
the case of central
collisions, after rescaling the averages with the 
corresponding value of the overlap function.

The simple picture of the interaction 
just described is easily tested experimentally. The universal
behaviour of the distributions in the number of observed jets,
just discussed, can be tested in a model independent way, which allows
a rather strict and detailed control on the different theoretical 
elements which enter in the representation
of the process.
Once the production mechanism is tested in detail, 
possible deviations from the universal behavior in particular
conditions (heavy nuclei, high energy, central collisions) 
would represent 
a convincing signal of a different production
mechanism taking place. 

\vskip.25in
{\bf Acknowledgements}

\vskip.25in

This work was partially supported by the Italian Ministry of University and of
Scientific and Technological Research by means of the Fondi per la Ricerca
scientifica - Universit\`a di Trieste.

\newpage

\begin{center}
{\bf Figure captions}\\
\end{center}
\vskip.15in
Fig. 1{\it a}. Projection in transverse plane of the 
scattering centres in a collision of a
parton with a $Pb$ nucleus. The scale is the $fm$ and the size of
the dots corresponds to the parton-parton cross section integrated 
with the cut off in momentum transfer $k_t^c=5GeV$. The momentum
fraction $x_A$ of the projectile parton is
$x_A=10^{-3}$ and the energy in the nucleon-nucleon c.m. system
is $7TeV$.
\vskip.15in
Fig. 1{\it b}. Same as in fig. 1{\it a} with $x_A=10^{-1}$.
\vskip.15in
Fig. 2{\it a}. Rapidity distribution of produced jets in a central 
$Pb-Pb$ collision with $7TeV$ energy in the 
nucleon-nucleon c.m. system. The histogram is the simulated
event, the continuous line is the average number of wounded partons
expected, as from Eq.(7), summing the projectile and target wouded 
partons
and expressing the resulting numbers as a function of rapidity. The dashed line
is
the same average number when using the approximate expression in Eq.(9),
namely after neglecting semi-hard parton rescatterings. $k_t^c=20GeV$.
\vskip.15in
Fig. 2{\it b}. Same as in fig. 2{\it a} with $k_t^c=10GeV$.
\vskip.15in
Fig. 2{\it c}. Same as in fig. 2{\it a} with $k_t^c=5GeV$.
\vskip.15in
Fig. 3{\it a}. Distribution in the total number of jets produced 
in central collisions with cutoff $k_t^c=20GeV$.
\vskip.15in
Fig. 3{\it b}. Same as in fig. 3{\it a} with $k_t^c=10GeV$.
\vskip.15in
Fig. 3{\it c}. Same as in fig. 3{\it a} with $k_t^c=7GeV$.
\vskip.15in
Fig. 4{\it a}. Relative rates of events with jets at $y=0$, $k_t^c=20GeV$
in the case of a Woods-Saxon distribution 
in space of the nuclear partonic matter 
(continuous histogram)
and for an hard-sphere distribution (dottet histogram).
\vskip.15in
Fig. 4{\it b}. Ratio of the two histograms in fig. 4{\it a}.

\end{document}